\documentclass[twocolumn,letterpaper]{IEEEAerospaceCLS}  

\usepackage{hyperref}
\usepackage[]{amsmath}    
\usepackage[]{graphicx}    
\newcommand{\ignore}[1]{}  

\begin{document}
\title{ Target Detection on Hyperspectral Images Using MCMC and VI Trained Bayesian Neural Networks }

\author{%
Daniel Ries\\ 
Sandia National Laboratories\\
PO Box 5800 MS 0829 \\
Albuquerque, NM 87185-0829 \\
dries@sandia.gov
\and 
Jason Adams\\ 
Sandia National Laboratories\\
Albuquerque, NM 87185\\
jradams@sandia.gov
\and
Joshua Zollweg\\ 
Sandia National Laboratories \\
Albuquerque, NM 87185 \\
jdzollw@sandia.gov
\thanks{\footnotesize \copyright2022 IEEE. Personal use of this material is permitted.  Permission from IEEE must be obtained for all other uses, in any current or future media, including reprinting/republishing this material for advertising or promotional purposes, creating new collective works, for resale or redistribution to servers or lists, or reuse of any copyrighted component of this work in other works. This is the accepted version of the article published available at \url{https://doi.org/10.1109/AERO53065.2022.9843344}}              
}

\maketitle

\thispagestyle{plain}
\pagestyle{plain}

\maketitle

\thispagestyle{plain}
\pagestyle{plain}

\begin{abstract}
Neural networks (NN) have become almost ubiquitous with image classification, but in their standard form produce point estimates, with no measure of confidence. Bayesian neural networks (BNN) provide uncertainty quantification (UQ) for NN predictions and estimates through the posterior distribution. As NN are applied in more high-consequence applications, UQ is becoming a requirement. Automating systems can save time and money, but only if the operator can trust what the system outputs. BNN provide a solution to this problem by not only giving accurate predictions and estimates, but also an interval that includes reasonable values within a desired probability. Despite their positive attributes, BNN are notoriously difficult and time consuming to train. Traditional Bayesian methods use Markov Chain Monte Carlo (MCMC), but this is often brushed aside as being too slow. The most common method is variational inference (VI) due to its fast computation, but there are multiple concerns with its efficacy. MCMC is the gold standard and given enough time, will produce the correct result. VI, alternatively, is an approximation that converges asymptotically. Unfortunately (or fortunately), high consequence problems often do not live in the land of asymtopia so  solutions like MCMC are preferable to approximations.

 We apply and compare MCMC- and VI-trained BNN in the context of target detection in hyperspectral imagery (HSI), where materials of interest can be identified by their unique spectral signature. This is a challenging field, due to the numerous permuting effects practical collection of HSI has on measured spectra. Both models are trained using out-of-the-box tools on a high fidelity HSI target detection scene.  Both MCMC- and VI-trained BNN perform well overall at target detection on a simulated HSI scene. Splitting the test set predictions into two classes, high confidence and low confidence predictions, presents a path to automation. For the MCMC-trained BNN, the high confidence predictions have a 0.95 probability of detection with a false alarm rate of 0.05 when considering pixels with target abundance of 0.2. VI-trained BNN have a 0.25 probability of detection for the same, but its performance on high confidence sets matched MCMC for abundances $>$0.4. However, the VI-trained BNN on this scene required significant expert tuning to get these results while MCMC worked immediately. On neither scene was MCMC prohibitively time consuming, as is often assumed, but the networks we used were relatively small. This paper provides an example of how to utilize the benefits of UQ, but also to increase awareness that different training methods can give different results for the same model. If sufficient computational resources are available, the best approach rather than the fastest or most efficient should be used, especially for high consequence problems. 
\end{abstract}

\tableofcontents

\section{Introduction}
\begin{figure}[h]
\centering
\includegraphics[width=0.3\textwidth]{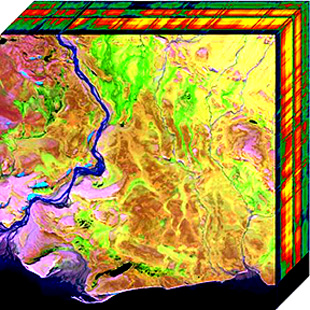}
\caption{Example of a hyperspectral image cube. Spatial coordinates are shown in the X/Y plane while the spectral coordinate is the Z plane. Image credit: \url{https://en.wikipedia.org/wiki/Hyperspectral_imaging}.}
\label{hsi}
\end{figure}

Aerial and aerospace assets collect imaging data in a variety of forms, and the remote detection of trace, sub-pixel targets in that image data is an important topic for a variety of applications. Hyperspectral imagery (HSI) contain hundreds of contiguous spectral bands which provide powerful information to detect material that would otherwise be near impossible. Aerial sensors with HSI measuring capabilities collect data that looks like  Figure \ref{hsi}, the three-dimensional cube containing both spatial information and spectral information. Target detection using HSI is a research area which has received significant attention in recent years \cite{nasrabadi2013,poojary2015,anderson2019}, and results have shown it is effective at finding rare targets \cite{anderson2019}. Uncertainty quantification (UQ) of model predictions is becoming a necessity in high consequence problems \cite{trucano2004,begoli2019}.

Aerospace sensors searching for targets of interest often use automated algorithms to make detections. Simple detectors such as the adaptive cosine estimator have proved time and again to give strong results \cite{manolakis2013}, but in the era of machine learning (ML) and artificial intelligence (AI), the common thought is that more advanced algorithms and detectors ought to provide better performance and generalization. However, traditional ML and AI methods only provide a best estimate, and do not provide an estimate of the model's confidence in oneself. This can be problematic for many high-risk aerospace target detection applications. 

Bayesian neural networks (BNN) were first popularized by David MacKay \cite{mackay1992,mackay1995} and his student Radford Neal \cite{neal1995,neal1996}. Neal's dissertation introduced Hamiltonian Monte Carlo (HMC) to sample the posterior distribution of a BNN, providing a practical way of training. To this day, HMC is considered the gold standard for BNN training due to its theoretical backing and lack of approximations. Before HMC, Gaussian approximations were typically used \cite{mackay1992,mackay1995}. \cite{muller1998} followed this up with alternative Markov Chain Monte Carlo (MCMC) methods for fixed architectures and \cite{insua1998} proposed an approach which treated the model architecture as unknown and estimated its posterior distribution with reversible jump MCMC (RJMCMC). \cite{andrieu2000} extended this work on RJMCMC. 

There were early applications of BNN in the statistics literature, including in time series \cite{liang2005}, medicine \cite{chakraborty2005,liang2018}, and with count data \cite{rodrigo2019}. \cite{titterington2004} provides a review of BNN and their common estimation methods at the time, MCMC, Gaussian approximation, and early variational inference (VI), from a statistical perspective.

Due to the increasing size of network architectures and the associated computational costs, faster sampling or approximation methods to obtain posterior distributions were explored. \cite{chen2014} introduced stochastic gradient HMC which uses a noisy estimate of the gradient from a subset of the data instead of the exact computation using all the data. \cite{li2019} extends this by applying variance reduction tricks which help speed convergence. 

Variational inference (VI) is the most popular method of Bayesian inference for NN \cite{graves2011}. \cite{blei2017} gives an extensive review of VI methods.  \cite{blundell2015} introduced Bayes by Backprop which is a practical stochastic VI algorithm to train a BNN. A common criticism of standard implementations of VI is the mean-field assumption, or assuming posterior independence of all parameters. \cite{louizos2016,zhang2017} each proposed new approaches to VI which allowed for training of full covariance variational distributions.   \cite{hernandez2015} introduced probabilistic backpropagation for scalable learning. Wang and Blei (2019) \cite{wang2019} established the frequentist consistency properties of VI, including asymptotic posterior Normality and consistency and asymptotic Normality of the posterior VI expectation, establishing VI as a serious large-sample alternative to MCMC. 

Although the introduction of new methods to provide UQ in deep learning is popular, there is less focus on ensuring the UQ provided by these methods is useful and transparent.  \cite{yao2019} compared UQ performance using various BNN training methods and using various metrics. The authors concluded a new metric for assessing predictive uncertainty is needed.\cite{wenzel2020} argue using various performance metrics that standard BNN can perform poorly with respect to UQ and propose using temperature scaling, otherwise known as weighted likelihood to make training adjustments. 

In this paper, we explore the performance of two different estimation methods for Bayesian inference and prediction. Although both methods will give the same results asymptotically under mild conditions, it is not always clear how fast asymptopia arrives nor do many applications in aerospace typically have large numbers of (labeled) observations of targets of interest. The quality of approximation for MCMC is determined by computer run time, or how long the MCMC sampler is run, while the quality of approximation for VI is determined by the data sample size. In a data poor environment, the ability of VI to produce similar results to MCMC needs to be assessed. By estimating  the same BNN  with MCMC and VI with the same training data, we will evaluate the relative performance of each. We compare MCMC on two data sets, a simple simulated regression problem and a high fidelity simulated HSI target detection problem.

This paper is organized as follows. In Section 2, the high fidelity simulated HSI scene, Megascene, is described as well as what the targets are and how they were added to the scene. In Section 3, the model and model fitting details are explained. Results are presented in Section 4 regarding predictive power and its intersection with UQ.  Section 5 summarizes our conclusions and discusses implications and future research directions.

\section{Data}

\begin{figure}[h]
\centering
\includegraphics[width=0.3\textwidth]{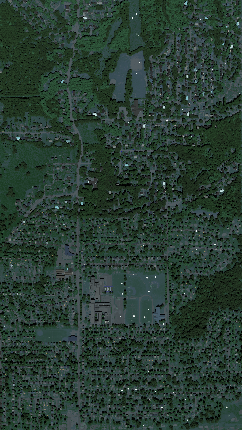}
\caption{Pseudo color render of Megascene MLS 1200.}
\label{scene}
\end{figure}

In order to have a scene for which we know ground truth and that represents our problem, we opted to create a synthetic dataset from DIRSIG Megascene \cite{ientilucci2003}. Megascene is modeled after a section of Rochester, NY and contains manmade objects such as houses and roads as well as natural features such as trees and grass. The simulator uses an AVIRIS-like sensor measuring 211 spectral bands ranging from 0.4 to 2.5 $\mu m$, creating a datacube similar to Figure \ref{hsi}. The images were created over the scene at an elevation of 4 km which gives a pixel size of 1 m$^2$.  A total of nine images were generated across three atmospheres (mid-latitude summer (MLS), sub artic summer (SAS), tropical (TROP)) and three times of day (1200,1430,1545). Figure \ref{scene} shows a pseudo color rendering of MLS 1200.

\begin{figure}[h]
\includegraphics[width=0.5\textwidth]{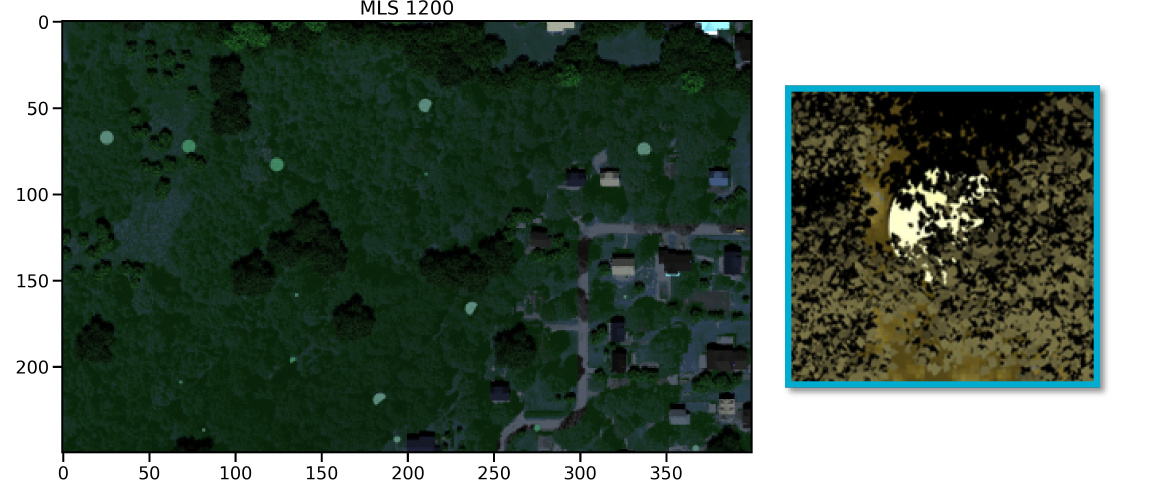}
\caption{Close up of several different sized target green discs in Megascene. On the right is a further zoomed in picture of one of the target discs that is partially hidden by foliage. Image credit to \protect\cite{anderson2019}. }
\label{targets}
\end{figure}

To serve as targets, we manually inserted green discs randomly through each scene. Each scene had 125 discs ranging in size from 0.1 to 4m radii, meaning some targets filled multiple pixels while others filled a small fraction of a pixel. A subset of the discs was made such that they were partially hidden beneath foliage, so not all the targets were complete circles. Figure \ref{targets} (Figure 6 in \cite{anderson2019}), shows an example of several different sized green target discs placed in Megascene. Some of the target discs were placed under foliage, as shown on the right image.

\begin{figure}[h]
\includegraphics[width=0.5\textwidth]{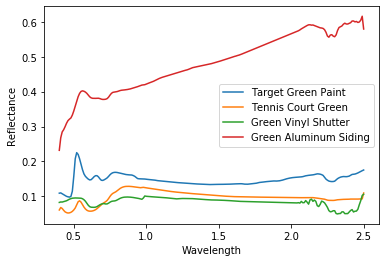}
\caption{Spectra of different green objects used in the creation of Megascene.}
\label{spectra}
\end{figure}

Figure \ref{spectra} shows the spectra for several different green objects used to create Megascene, these are common confusers for our green paint target. Most pixels in the dataset will be a combination, or mixture, of several materials' spectra since with a pixel size of 1m$^2$, there is often more than one material in the area.

The left half of MLS-1200 was used for training the BNN models. The right half of all nine scenes were used as test sets. By only training on one scene at a particular time and atmosphere, we are able to understand the model's ability to detect targets in scenes it has never seen before. This is particularly important for our application since we cannot expect to have training data in all atmospheres and times of day due to expense and practicality reasons. Even though aerospace sensors might be able to collect data at many different atmospheres and times of day, it is costly to have labeled data at all these combinations. 

\section{Methods}

The BNN contained 3 hidden layers, each with 10 neurons activated with a sigmoid function. Although ReLU tends to be more computationally efficient we found sigmoid to give better results. The priors on the weights were all Normal with mean 0 and standard deviation 10. The standard deviation was selected partially to optimize performance, making this a quasi-Empirical Bayes approach. Although we do not believe this is the best approach, there are others working on BNN priors \cite{ghosh2018} and this serves as a proof of concept.

Because the features are spectra, they are functional data by nature and contain a correlated structure. The structure has physical meaning itself and can be used for model explainability  \cite{goode2020}. To account for this dependence and help reduce the dimensionality of the inputs, we employ functional principal component analysis (fPCA) on the feature functions. We then use the first 25 functional principal components (fPC) for each pixel as features. The first 25 fPCs explain 99.999\% of the variability, and later fPCs did not add to the predictive power of the model. Formally, the model is written as:

\begin{align}
Y_i &\overset{iid}{\sim} Bernoulli(\pi_i), i=1,2,...,n \\
\pi_i &= f(\boldsymbol{\theta}, {\bf x}_i) \\
\theta_j &\overset{iid}{\sim} N(0,10), j=1,2,...J
\end{align}

where $Y_i$ is a binary random variable for whether pixel $i$ contains target with parameter $\pi_i \equiv P(Y_i = 1)$, the probability that pixel $i$ contains target.  This is itself a deterministic function of the 25 fPCs for pixel $i$, ${\bf x}_i$, the NN model $f(\cdot,\cdot)$, and the NN's parameters $\boldsymbol{\theta}$. Note $\pi_i = \pi_i({\bf x_i},\boldsymbol{\theta})$, but we drop the dependence for brevity. Because the probability $\pi_i$ is an unknown parameter, it is treated as a distribution in Bayesian statistics, with its prior distribution inferred by the prior on $\boldsymbol{\theta}$. Denote the vectors ${\bf Y}=(Y_1,Y_2,...,Y_n)$ and $\boldsymbol{\theta} = (\theta_1,\theta_2,...,\theta_J)$.

All pixels which had a target abundance greater than zero were used for training, and about ten times the number of pixels with target abundance zero were randomly sampled and used for training. This subsetting sped up training significantly due to the sparse nature of the targets in the scene. Reducing the number of non-targets did not affect the model's performance. We use numpyro and pyro to fit the BNN via MCMC and VI, respectively \cite{phan2019,bingham2019}. 

BNN output a posterior distribution for $\pi_i$, denoted $p(\pi_i|{\bf Y})$. A posterior mean can then be used as a point estimate by taking $E(p(\pi_i|{\bf Y}))$. Uncertainty around $\pi_i$ can be quantified using confidence intervals (sometimes called credible intervals.) These intervals are constructed by taking the $\alpha/2$th and $(1-\alpha/2)$th quantiles of $p(\pi_i|{\bf Y})$ , denoted as $L_{\alpha,i}, U_{\alpha,i}$, respectively, to create a $1-\alpha$ confidence interval. Based on Bayesian probability, $P(L_{\alpha,i} < \pi_i < U_{\alpha,i}|{\bf Y}) = 1-\alpha$. The value of $\alpha$ is chosen depending on the risk desired.

One way to incorporate the UQ provided by the BNN is using high confidence (HC) sets. An HC set contains predictions that are either close to 0 or 1 indicating a high probability of either no-target or target and with a corresponding confidence interval that spans no more than a specified width. More formally,  pixel $i$ is included in the HC set $\Omega$:

\begin{align}
    i \in \Omega \iff P(\pi_i < \text{$\mathcal{L}$}|{\bf Y}) > 1-\alpha  \text{ OR }  P(\pi_i > \text{$\mathcal{U}$}|{\bf Y}) > 1-\alpha
    \label{def:hc}
\end{align}

where $\mathcal{L}$ ($\mathcal{U}$) is the value the estimated target probability for pixel $i$ ($\pi_i$)  needs to be less (greater) than, and $1-\alpha$ is the desired confidence that $\pi_i$ is less (greater) than $\mathcal{L}$ ($\mathcal{U}$). Therefore, if pixel $i$ is in the high confidence set, we can say, there's at least a $1-\alpha$ probability that $\pi_i$ is less than $\mathcal{L}$ (greater than $\mathcal{U}$). This ensures two things: (i) the estimated probability of pixel $i$ containing target is either close to 0 or close to 1, as defined by chosen $\mathcal{L}$ and $\mathcal{U}$, and (ii) we are confident in the estimated probability of pixel $i$ containing target since there's at least a $1-\alpha$ chance that $\pi_i$ is less than $\mathcal{L}$ (greater than $\mathcal{U}$). For this application we choose $\mathcal{L}$ $ =0.2, \mathcal{U}$ $ =0.8, \alpha=0.2$.  This set contains predictions which are strongly target or non-target and the model has high confidence in that prediction.

\section{Results}

The MCMC model ran on 2 chains for 2500 iterations with 500 burn in period. The chains were run in parallel and total MCMC training time was about 21 minutes on a Intel(R) Xeon(R) CPU E5-2650 v4 2.20GHz. Posteriors estimated from MCMC must be checked for proper convergence, but overparameterized BNN weights may not be identifiable. Checking convergence of predictions across chains is the alternative to ensure the MCMC behaves accordingly. We checked many pixels' prediction traces and there were no signs of non-convergence meaning the MCMC is behaving as expected. VI was optimized using Adam with a learning rate of 0.01 for 450 epochs, monitoring the validation loss for overfitting. The VI training was much faster, taking only about 4 seconds on the same CPU. The VI model could have been trained using a GPU, which often trains faster than CPU, but given the relatively simple architecture it was unnecessary to do so. In this situation, 21 minutes is not cost prohibitive for a real life target detection algorithm, so the difference in computation time is of minor concern compared to model performance. 

\begin{figure}[h]
\includegraphics[width=0.5\textwidth]{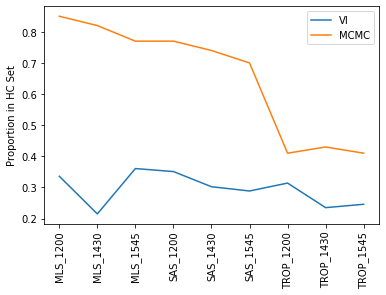}
\caption{Proportion of pixels in high confidence set across nine tests scenes for MCMC- and VI- trained BNN.}
\label{hc_proportion}
\end{figure}

 Figure \ref{hc_proportion} shows the proportion of data in the HC set for each scene for both the MCMC- and VI-trained models. Overall, the MCMC model creates larger HC sets.  For MLS and SAS scenes the MCMC-trained model contains over twice as many pixels as VI. The MCMC model has a large drop in HC set pixels for TROP, but still more than VI.  VI is fairly constant across scenes.  This is interesting since theory tells us uncertainties  from mean-field VI should under represent the true uncertainties due to independence assumptions on the posterior, and although this result is not exactly a test of that theory, it is unexpected that the VI model appears to have refrained from being overly confident. However, note that the proportion of data within the HC set is not an evaluation of the model, rather an outcome. That is, pixels are included or excluded based on the UQ given by the model, therefore an overly conservative model may not include all pixels that are truly high confidence or an overly optimistic model may include pixels that have no business being called highly confident. Future work needs to address the quality of UQ given by models to ensure membership to HC sets retains its quantitative meaning. 

\begin{figure}[h]
\includegraphics[width=0.5\textwidth]{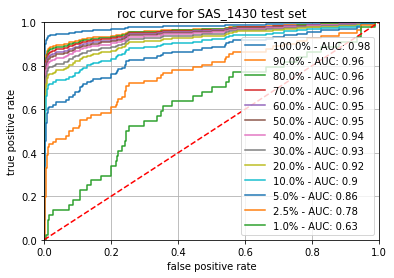}
\caption{ROC evaluated on FULL SAS 1430 test scene for MCMC-trained BNN. Lines denote ROC scores for the sets of pixels containing target in proportions up to the denoted fraction. Area under the curves are given next to each fraction.}
\label{mcmc_full_roc}
\end{figure}

\begin{figure}[h]
\includegraphics[width=0.5\textwidth]{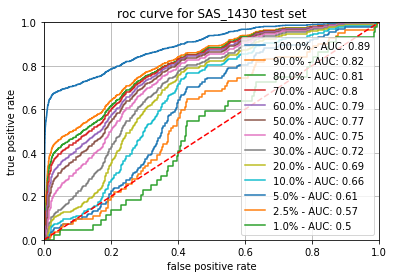}
\caption{ROC evaluated on FULL SAS 1430 test scene for VI-trained BNN. Lines denote ROC scores for the sets of pixels containing target in proportions up to the denoted fraction. Area under the curves are given next to each fraction.}
\label{vi_full_roc}
\end{figure}

Figures \ref{mcmc_full_roc} and \ref{vi_full_roc} show ROC curves for MCMC- and VI-trained BNNs on the full  SAS 1430 scene, respectively. We only show ROC curves from one of the nine scenes, but the trends are the same for all scenes. The  lines denote ROC scores for the sets of pixels containing target in proportions up to the denoted fraction. This allows evaluation at different abundance levels to see how good the models are at finding different sized targets. This is important because depending on the resolution of the remote sensing device, a pixel could represent a relatively large area, and sub-pixel detection is necessary. Overall, the MCMC BNN has much better performance, often having area under the ROC curves about 10 percentage points higher. Additionally, the MCMC model tends to have much higher detection rates at low false alarm rates, which is important in high consequence national security problems. 

\begin{figure}[h]
\includegraphics[width=0.5\textwidth]{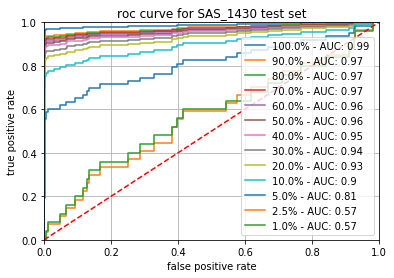}
\caption{ROC evaluated on HIGH CONFIDENCE SAS 1430 test scene for MCMC-trained BNN. Lines denote ROC scores for the sets of pixels containing target in proportions up to the denoted fraction. Area under the curves are given next to each fraction.}
\label{mcmc_hc_roc}
\end{figure}

\begin{figure}[h]
\includegraphics[width=0.5\textwidth]{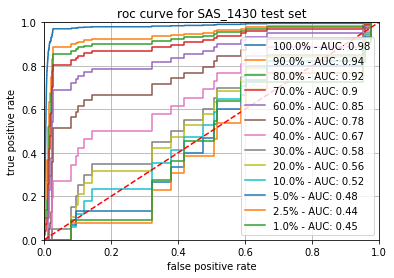}
\caption{ROC evaluated on HIGH CONFIDENCE SAS 1430 test scene for VI-trained BNN. Lines denote ROC scores for the sets of pixels containing target in proportions up to the denoted fraction. Area under the curves are given next to each fraction.}
\label{vi_hc_roc}
\end{figure}

Figures \ref{mcmc_hc_roc} and \ref{vi_hc_roc} show ROC curves for MCMC- and VI-trained BNNs on the HC set only on SAS 1430, respectively. These ROC curves show the performance when we only consider pixels for which the model determines it is confident in its prediction. In this case, there is a slight bump in performance for the MCMC model for abundances down to 20\%, and then there is actually a degradation in performance for abundances $<$10\%. This isn't surprising since the model is more likely to be confident for pixels which contain a high proportion of target compared to pixels containing a very small amount, where its target signature could be mixed with the background. The VI-BNN sees a large boost in performance for abundances $>$50\%, and slight degradations for abundances $<$50\%, likely for the same reasons. 

\begin{figure}[h]
\includegraphics[width=0.5\textwidth]{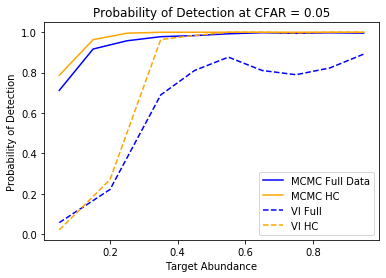}
\caption{Probability of detection as a function of target pixel abundance for a constant false alarm rate of 0.05. MCMC-trained BNN results are solid lines, VI-trained BNN results are in dashed lines. Full data test results are in blue, and high confidence set results are in orange.}
\label{cfar}
\end{figure}

For national security problems, low false alarm rates are often needed. Furthermore, we are most interested in detection at low pixel abundance levels. Therefore we assess model performance by looking at probability of detection as a function of target abundance, for a constant false alarm rate of 5\%.  Figure \ref{cfar} shows detection probabilities averaged over all nine scenes for both HC sets and full test sets, for the MCMC- and VI-trained BNN. The MCMC BNN performs much better at low pixel target abundances compared to VI. This effect diminishes somewhat at an abundance level of 40\%. HC sets also provide a boost for both MCMC and VI methods, but at different times. Although the BNN model is trained in two different ways, we should still expect results from the same model to only differ slightly based on training approach, and these prediction results show the converse is true. One consideration is more tuning could be done with the VI algorithm to improve optimization and thus prediction, whereas tuning the MCMC was fairly straightforward for this application. 

Another common misconception is the predicted probability of target, or output of a standard NN classifier, fully quantifies the uncertainty of the prediction. This predicted probability only contains the aleatoric uncertainty as determined by the variance of a Bernoulli random variable (in the case of target detection). This prediction does not contain the epistemic uncertainty, part of which is the modeling and sampling uncertainty. A predicted target probability of 0.99 does not automatically imply we should be confident in the prediction itself, rather that is the model's best guess if it was forced to predict. The confidence interval from the posterior determines the model's confidence in the prediction, and intervals are not always symmetric. It is possible for a prediction of 0.99 to have a 90\% confidence interval ranging from 0.01 to 0.999. In high consequence national security problems, decisions should not be made solely with the point estimate when we can know the model's confidence in the prediction. It turns out, this is not an unrealistic scenario. 

\begin{figure}[h]
\includegraphics[width=0.5\textwidth]{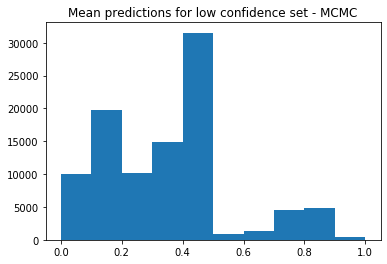}
\caption{Distribution of low confidence set predictions from MCMC-BNN on the nine combined test sets for pixels containing target.}
\label{low}
\end{figure}

\begin{figure}[h]
\includegraphics[width=0.5\textwidth]{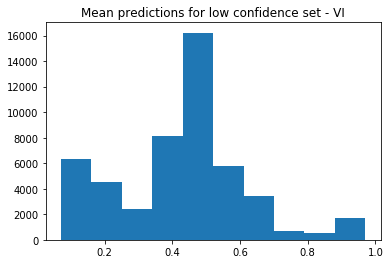}
\caption{Distribution of low confidence set predictions from VI-BNN on the nine combined test sets for pixels containing target.}
\label{low_vi}
\end{figure}

Figure \ref{low} and \ref{low_vi} show the distribution of low confidence (LC) predictions on the nine combined test sets for pixels containing target, for MCMC and VI, respectively. The LC set is the opposite of the HC set, that is it contains predictions whose lower confidence bound is less than 0.2 and upper confidence bound is greater than 0.8. As expected, most point predictions are close to 0.5, but many are towards 0 and 1. Although these represent a relatively small number of the overall predictions, they are not negligible in high consequence situations where every false positive or false negative can be extremely costly. The fact that there are target probability estimates close to 0 or 1 that are in the low confidence set further confirms that having uncertainty quantification on estimates is imperative to avoid a false degree of confidence.  Predictions which fall in the low confidence set should receive further review because the model is incapable of providing information on them, and simply having an estimate close to 0 or 1 is not evidence enough to make assumptions about them.

\begin{figure*}[h]
\centering
\includegraphics[width=\textwidth]{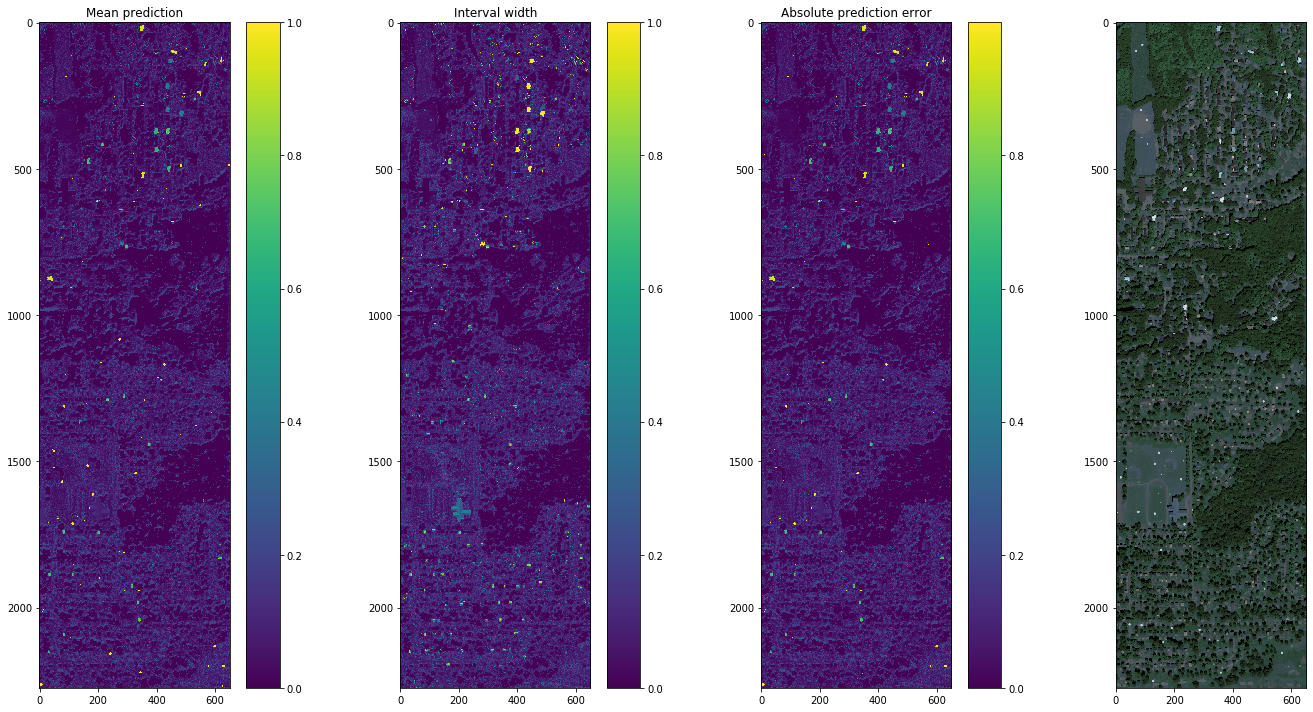}
\caption{MLS 1200 test set mean prediction, interval width, absolute prediction error, and rgb image, respectively for MCMC-trained BNN.}
\label{scene_heatmap}
\end{figure*}

Figure \ref{scene_heatmap} shows the MLS 1200 test set mean prediction, interval width, absolute prediction error, and rgb image, respectively for MCMC-trained BNN. These plots are useful to understand where the model predicts targets, where the model is confident in its predictions, and where the model makes mistakes. Looking at the rgb image, the context and spatial surroundings of the problem can be identified, and further examination of the spectra can indicate what types of materials confuse the model or cause the model to be uncertain. Knowledge of these shortcomings can then be used to efficiently design future data collection campaigns and be communicated to operators, so they understand the model's shortcomings. 

BNN are able to perform target detection on this scene and utilize their uncertainties to reduce false alarms. Although the underlying BNN model is the same, the training method differs and gives surprisingly different results. As the gold standard, the MCMC results show the true power of the BNN, especially when considering high confidence sets. The VI results show promise to the posterior approximation method and present a computationally efficient alternative. The out-of-the-box approach for VI requires expert knowledge to understand, set up the model, and tune the hyperparameters. Both models show how UQ can be used in practice to handle high consequence problems and both showed the limitations of using estimated class probabilities themselves as model uncertainty.

\section{Discussion}

In this paper we compared the results in predictive performance and uncertainty quantification for target detection in a HSI problem  using BNNs, trained using both MCMC and VI. MCMC is generally considered the gold standard to compare Bayesian model results to, and that held in this experiment as well. Results from the VI model were slightly worse than the MCMC model and required more effort tuning. Given that NN are commonly used ``off-the-shelf" already by practitioners, it is only a matter of time before BNN are an off-the-shelf tool to provide uncertainty quantification, and VI is the obvious computationally efficient tool to provide that training. This paper is meant to provide an example of how to utilize the benefits of UQ, but also to increase awareness that the optimization method matters, and if sufficient computational resources are available, the best approach rather than the fastest or most efficient should be used, especially for high consequence problems. 

This paper shows there is still work to be done with all-purpose, generic VI algorithms before they can be used by non-deep learning experts. Furthermore, we gave examples of why uncertainty quantification of estimates and predictions is important in practice, specifically in classification problems when researchers and practitioners alike often interpret the estimated class probability as the uncertainty in that same prediction. 

These results and insights are relevant to the aerospace community because remote sensing assets collect a wealth of HSI information that often needs to be analyzed quickly and accurately. For remote sensing in the national security space, having a reliable model that is confident in its predictions is imperative due to the high consequence nature of the decisions. BNN show promise for providing this capability, and while MCMC provides quality predictions and uncertainty, it is often computationally infeasible for most problems. Progress is being made on  more efficient, available off-the-shelf VI, but there still exist limitations for this method to be utilized without a deep learning expert.

\acknowledgments
Sandia National Laboratories is a multimission laboratory managed and operated by National Technology \& Engineering Solutions of Sandia, LLC, a wholly owned subsidiary of Honeywell International Inc., for the U.S. Department of Energy's National Nuclear Security Administration under contract DE-NA0003525. This paper describes objective technical results and analysis. Any subjective views or opinions that might be expressed in the paper do not necessarily represent the views of the U.S. Department of Energy or the United States Government. SAND2021-15821 C.

\bibliographystyle{IEEEtran}

\bibliography{bibliography}

\thebiography

\begin{biographywithpic}
{Daniel Ries}{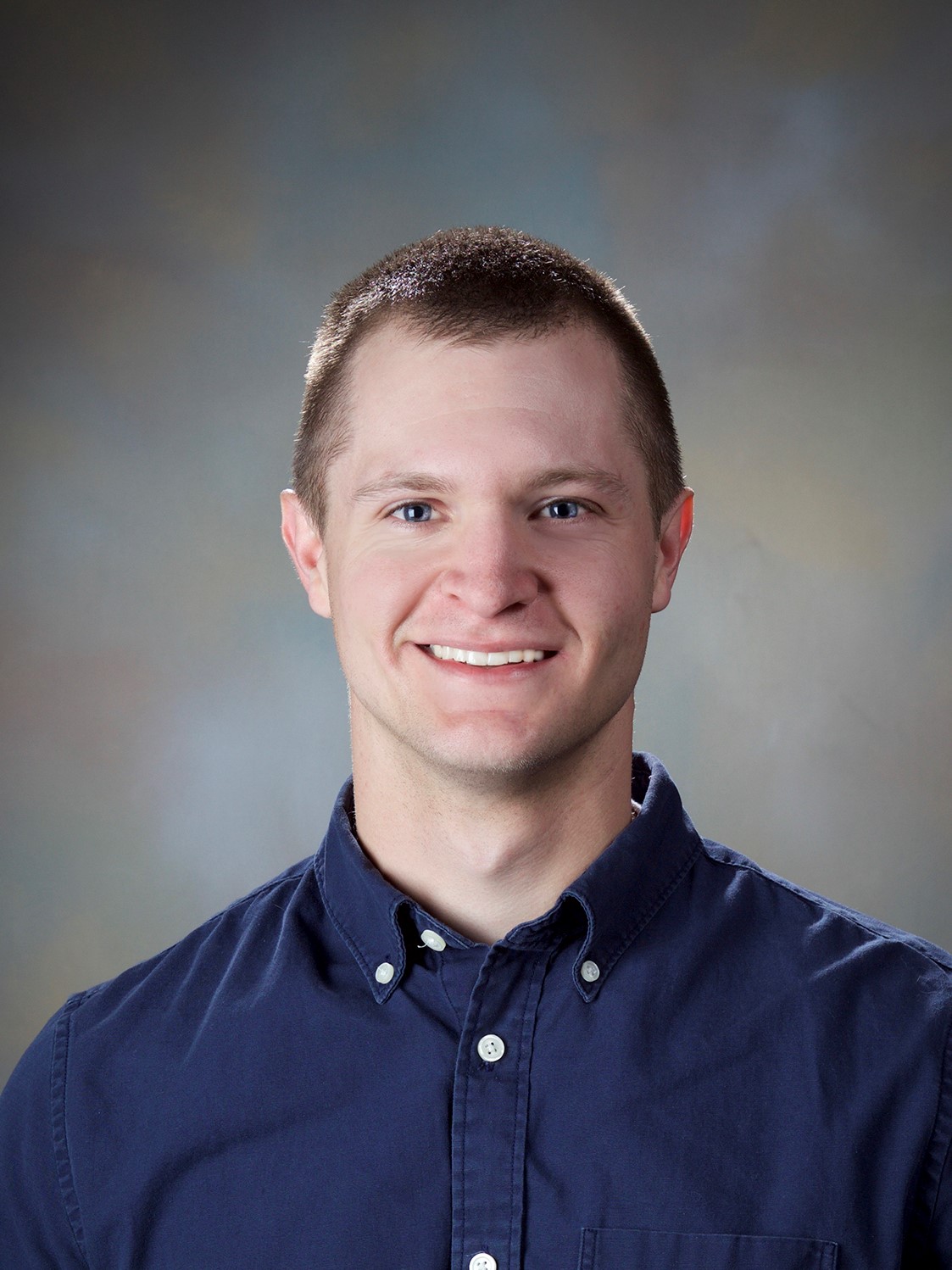}
received his BS degree in Economics from the University of Minnesota, and his MS and PhD degrees in statistics from Iowa State University. He is currently a Senior Member of the Technical Staff in the Statistics and Data Analytics Department at Sandia National Laboratories. His research interests include uncertainty quantification for deep learning models, functional data analysis, and Bayesian modeling.
\end{biographywithpic} 

\begin{biographywithpic}
{Jason Adams}{JAdams}
is a senior member of the technical staff at Sandia National Laboratories. He received a PhD in Statistics from the University of Nebraska-Lincoln. His research interests include incorporating uncertainty quantification in deep learning models, machine learning for image analysis, functional data analysis, and computational statistics.
\end{biographywithpic}

\begin{biographywithpic}
{Joshua Zollweg}{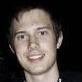}
 holds BS and MS degrees (2010, 2012) in Imaging Science from the Rochester Institute of Technology. He was a Principal Member of Technical Staff at Sandia National Laboratories until 2021, where he has started in 2012. Joshua's work at Sandia centered around simulating remotely sensed data and developing target detection algorithms. 

\end{biographywithpic}

\end{document}